\documentstyle[aps,multicol,pre,epsfig]{revtex}

\newcommand{\beq}{\begin{equation}}
\newcommand{\eeq}{\end{equation}}

\begin{document}

\begin{title}
{\bf Dynamics of shallow dark solitons in a trapped gas of impenetrable bosons}
\end{title}

\author{
D.J.\ Frantzeskakis$^{1}$,  N.P.\ Proukakis$^{2}$ and P.G.\ Kevrekidis$^{3}$}
\address{
$^{1}$ Department of Physics, University of Athens, Panepistimiopolis,
Zografos, Athens 157 84, Greece \\
$^{2}$ Department of Physics, University of Durham, South Road, Durham DH1 3LE, 
United Kingdom \\
$^{3}$ Department of Mathematics and Statistics,
University of Massachusetts, Amherst MA 01003-4515, USA \\
}

\maketitle

\vspace{2mm}

\begin{abstract}
The dynamics of linear and nonlinear excitations in a 
Bose gas in the Tonks-Girardeau (TG) regime with longitudinal confinement  are 
studied within a mean field theory of quintic nonlinearity.
A reductive 
perturbation method is
used to demonstrate that the dynamics of 
shallow dark solitons, in the presence of an external potential, 
can effectively be described
by a variable-coefficient Korteweg-de Vries equation. The soliton oscillation frequency 
is analytically obtained to be equal to the axial trap frequency, in agreement with numerical predictions 
obtained by Busch {\it et al.} [J. Phys. B {\bf 36}, 2553 (2003)] via the Bose-Fermi mapping. 
We obtain  
analytical expressions for the evolution of both soliton and emitted radiation (sound) profiles. 
\end{abstract}


\begin{multicols}{2}


Dark solitons are fundamental nonlinear excitations of the nonlinear
Schr\"{o}dinger (NLS) equation  with repulsive interactions, which have 
attracted much attention in the field of atomic Bose-Einstein condensates 
(BECs), where they have been studied experimentally 
\cite{dark} and theoretically in the framework of the Gross-Pitaevskii (GP) 
equation  
\cite{motion,huang,frantz,dis,thsnbec,nick,nick2,george}.  
In particular, under conditions of tight transverse confinement, for which the 
gas becomes quasi-one-dimensional, dark solitons in 
longitudinal harmonic traps have been shown to oscillate with
frequency $\Omega/\sqrt{2}$, where $\Omega$ is the axial trap
frequency \cite{motion,huang,frantz}. Thermal \cite{dis} and
dynamical \cite{thsnbec} instabilities, including  
sound emission phenomena \cite{nick,nick2}, have also been
investigated. Furthermore, apart from the traditional rectilinear
solitons, ring-shaped dark solitons, have also been proposed
\cite{george}.

For very tight transverse confinement and sufficiently low densities ($n\ll 1/\xi$, where 
$\xi$ the healing length of the system) \cite{ols1}, an atomic gas with repulsive interactions 
enters the Tonks-Girardeau (TG) regime\cite{tg}, where it behaves like a one-dimensional gas of impenetrable bosons. This regime, which is currently under intense experimental investigation 
\cite{Tonks_Exp}, has attracted considerable theoretical attention both in the absence 
\cite{LL,gw,gw3}, as well as in the presence of axial confinement \cite{petrov,dunjko,santos}.
Dark solitons have also been predicted to occur in the TG regime, either upon employing the Bose-Fermi mapping theorem \cite{gw,busch}, or via a mean-field approach of quintic
nonlinearity arising in this limit \cite{kolom1,bhad} (see also relevant work for fermionic systems \cite{fermi}). However, some important issues of soliton dynamics, such as its instability to sound emission due to the longitudinal confinement, and the profiles of the emitted sound waves have not been discussed in detail. It is the aim of this paper to address these points {\it analytically} in the case of {\em shallow} solitons.

Our analysis is based on the modified NLS equation with a quintic nonlinearity (as opposed to the cubic nonlinear 
GP equation valid in 3D systems \cite{review})
given by 
\begin{equation}
i\hbar \frac{ \partial \psi}{\partial t}=\left[-\frac{\hbar^{2}}{2m}
\frac{\partial^{2}}{\partial x^{2}} +
\frac{\pi^{2}\hbar^{2}}{2m}|\psi|^{4}+ \frac{1}{2} m \omega_{x}^{2} x^{2} 
\right]\psi.
\label{gpetg}
\end{equation}
Here $\psi$ is the order parameter of the system (normalized to the number of particles $N$), 
$m$ the atomic mass and $\omega_{x}$ the axial confining frequency. 
The above long wavelength equation \cite{kolom1}, originally derived in 
\cite{kolom2} from a renormalization group approach, has also been obtained by other techniques, such as (i) the Kohn-Sham density functional theory for bosons 
\cite{bhad,tanatar,kim}, (ii) an in-depth study of the energy dependence of the many-body effective interaction \cite{lee0}, (iii) a rigorous analysis of the many-body Schr{\"o}dinger equation \cite{lieb}. This equation is expected to be valid for weak density modulations, as long as the number of atoms exceeds a certain minimum value (typically much larger than 10), for which oscillations in the density profiles become essentially suppressed 
\cite{kolom1,lee0,anna}. Note that an alternative density-functional approach to deal with the 
excitations and dynamics in the TG limit has been recently proposed \cite{brand}.

The validity of Eq. (\ref{gpetg}) has been criticized 
in certain regimes where the outcome is sensitive to the exact phase of the order parameter, e.g., interference experiments on a torus, where effects beyond the realm 
of mean field theory arise, and for which Eq. (\ref{gpetg}) overestimates the coherence properties \cite{gw2}. The latter treatment, however, ignores longitudinal confinement which 
is the key source of the dynamics studied in this paper. An alternative exact treatment 
\cite{busch} including the harmonic confinement has predicted soliton oscillations at the trap 
frequency (this result was explicitly demonstrated numerically for deep dark solitons). 
This paper complements the result of \cite{busch}, in that the same oscillation frequency 
is obtained {\it analytically} 
for {\em shallow} dark solitons, for which the employed hydrodynamic approximation is expected to be valid. 

Solitons moving on a background density gradient are prone to sound emission 
\cite{motion,huang,frantz,nick2}, but are stabilized against decay in a single harmonic 
trap by the continuous sound emission--reabsorption cycles \cite{nick}. We derive  analytical results for both the soliton amplitude and speed as well as associated sound profiles.
Starting from Eq. (1), our analytical predictions are obtained by means of a reductive 
multi-scale perturbative method, which yields a variable coefficient Korteweg-de Vries (KdV) equation for the dynamics of shallow dark solitons. Note that the above quintic NLS equation is a model of mathematical interest in its own right, as ``traditional'' analytical techniques, such as the adiabatic perturbation theory for dark solitons in BECs \cite{frantz,george}, cannot be applied for such nonlinearities. However, a similar technique as the one presented here has been succesfully applied to 
quasi-one-dimensional condensates in the GP limit \cite{huang}.
To start our analysis, we express Eq. (\ref{gpetg}) in  dimensionless
form,
\begin{equation}
iu_{t}=-\frac{1}{2}u_{xx}+|u|^{4}u+V(x)u,
\label{gpe}
\end{equation}
where the subscripts denote partial differentiation, while the 
variables $t$, $x$ and the function $u$ are measured, 
respectively, in units of $m/(\hbar \pi^{2} n_{\rm o}^{2})$, $1/(\pi n_{\rm o})$ and 
$2^{1/4}\sqrt{n_{\rm o}}$ (where $n_{\rm o} \equiv|\psi_{\rm o}|^{2}$ the peak density of 
the gas). The external potential in Eq. (\ref{gpe}) is $V(x)=(1/2)(\pi 
n_{\rm o}a_{x})^{-4}x^{2}$, where $a_{x}\equiv \sqrt{\hbar/(m\omega_{x})}$ is the 
harmonic oscillator length in the axial direction. Taking into regard that the 
parameter $(\pi n_{\rm o} a_{x})^{-4}$ is apparently small, it is convenient to 
define the small parameter 
$\epsilon \equiv \Omega^{-2/3}(\pi n_{\rm o} a_{x})^{-4/3}$ 
[$\Omega$ is a parameter of order $O(1)$], which will be used in the 
perturbation analysis to be presented below. This way, the external potential 
takes the form 
$V(X)=(1/2)\Omega^{2}X^{2}$, i.e., it is a function of the slow variable 
$X=\epsilon^{3/2}x$,
while $\Omega$ 
expresses the strength of the magnetic trap, or the 
normalized axial trap frequency. 
By analogy to 
\cite{anna}, 
we introduce the Madelung transformation $u=\sqrt{n}\exp(i\phi)$ ($n$ and $\phi$ denote the density and phase respectively) to reduce Eq. (\ref{gpe}) to the following hydrodynamic equations, 
\begin{eqnarray}
&& n_{t}+(n\phi_{x})_{x}=0, \label{h1} \\
&& \phi_{t}+\frac{1}{2}\phi_{x}^{2}+n^{2}-\frac{1}{2}n^{-1/2}(n^{1/2})_{xx}
+V(X)=0, \label{h2}
\end{eqnarray}
which are similar to the ones that have been employed to discuss the crossover from TG to BEC regime \cite{dunjko,santos,chiara}. 
The ground state of the system may  be obtained upon assuming that the atomic  
velocity $v\equiv \phi_{x}=0$ (i.e., no flow in the system) and $\phi_{t}=-\mu_{0}$ 
(dimensionless chemical potential). Then, as Eq. (\ref{h1}) implies that $n=n_{0}$ is 
time-independent in the ground state, we assume that $n_{0}=n_{0}(X)$. As the 
quantum pressure term 
in Eq. (\ref{h2}) is of order $O(\epsilon^{3})$, to leading order in $\epsilon$ 
[to $O(1)$], 
we obtain 
\begin{equation}
n_{0}(X)=\sqrt{\mu_{0}-V(X)},
\label{TF}
\end{equation}
in the region where $\mu_{0}>V(X)$, and $n_{0}=0$ outside. Equation (\ref{TF}) 
gives the density profile in the so-called 
Thomas-Fermi (TF) approximation \cite{review}.
It follows from Eq. (\ref{TF}) that in the case of the harmonic trap the axial 
size of the gas is $2L$, where $L=\sqrt{2\mu_{0}}/\Omega$. We now consider the
propagation of small-amplitude linear excitations (e.g., sound waves) of the 
ground state, by seeking solutions of Eqs. (\ref{h1})-(\ref{h2}) of the form,
\begin{equation}
n=n_{0}(X)+\epsilon\tilde{n}(x,t), \,\,\ \phi=-\mu_{0}t+\epsilon 
\tilde{\phi}(x,t),
\label{sw}
\end{equation}
where the functions $\tilde{n}$ and $\tilde{\phi}$ describe the linear 
excitations. Substituting Eqs. (\ref{sw}) in Eqs. (\ref{h1})-(\ref{h2}), to order 
$O(1)$ we recover the TF approximation, while to order $O(\epsilon)$ we obtain 
the following system of linear equations,
\begin{equation}
\tilde{n}_{t}+n_{0}\tilde{\phi}_{xx}=0, \,\,\,\
\tilde{\phi}_t+2n_{0}\tilde{n}-\frac{1}{4n_{0}}\tilde{n}_{xx}=0.
\label{s1}
\end{equation}
The dispersion relation of the excitations can easily be obtained, upon 
considering plane-wave solutions of Eqs. (\ref{s1}) of the form 
$(\tilde{n}, \tilde{\phi})=
(\tilde{n}_{0}, \tilde{\phi}_{0})\exp[i(kx-\omega t)]$, where
$\tilde{n}_{0}$ and $\tilde{\phi}_{0}$ are independent of $x$ and $t$
(but may depend on $X$). 
This way, we readily obtain the equation,
\begin{equation}
\omega=\pm \sqrt{2n_{0}^{2}(X)k^{2}+\frac{1}{4}k^{4}}
\label{b}
\end{equation}
which is a Bogoliubov-type excitation spectrum, but with the excitation 
frequency $\omega$ being a function of the slow variable $X$. The speed of 
sound is local and is given by
\begin{equation}
c_{S}=s \sqrt{2}n_{0}(X),
\label{cs}
\end{equation}
where $s={\rm sign}(c_{S})=\pm 1$, i.e., the sound may propagate in two opposite 
directions. Note that the local character of the speed of sound is due to the 
presence of the external potential, which bears resemblance to the sound propagation 
in slowly-varying nonuniform media \cite{landau}.

We now employ the reductive perturbation method \cite{rpm,asano} (see also 
\cite{huang} for a relevant study in BECs) to examine 
the evolution of the {\it nonlinear} excitations (e.g., solitons), of the ground 
state. As 
Eqs. (\ref{h1})-(\ref{h2}) are inhomogeneous, 
we introduce a new slow time-variable, 
$T=\epsilon^{1/2}\left( t-\int^{x} C^{-1}(x')dx'\right)$, where $C$ is the 
(local) velocity of the nonlinear excitations, to be determined in a self-consistent manner. Also, we introduce the following asymptotic expansions for the 
functions $n$ and $\phi$,
\begin{eqnarray}
n&=&n_{0}(X)+ \epsilon n_{1}(X,T)+\epsilon^{2} n_{2}(X,T)+\cdots, \nonumber \\
\phi&=&-\mu_{0}t+ \epsilon^{1/2} \phi_{1}(X,T)+\epsilon^{3/2} 
\phi_{2}(X,T)+\cdots.
\label{ae}
\end{eqnarray}
Substituting the expansions (\ref{ae}) into Eqs. (\ref{h1})-(\ref{h2}), we 
obtain the following results: First, to order $O(1)$, Eq. (\ref{h2}) leads to 
the TF approximation 
[see Eq. (\ref{TF})]. Then, to orders $O(\epsilon)$ and $O(\epsilon^{3/2})$, 
Eqs. (\ref{h2}) and (\ref{h1}) respectively lead to the following system of 
linear equations:
\begin{equation}
\phi_{1T}+2n_{0}n_{1}=0, \,\,\,\ n_{1T}+C^{-2}n_{0}\phi_{1TT}=0.
\label{linear}
\end{equation}
The compatibility condition of Eqs. (\ref{linear}) is the equation
$1-2n_{0}^{2}C^{-2}=0$, which determines the unknown 
velocity $C$, which, actually, is the same as the speed of sound, i.e.,
$C\equiv c_{S}$ [c.f. Eq. (\ref{cs})]. On the other hand, Eqs. (\ref{linear})
lead to the following equation,
\begin{equation}
\phi_{1}(X,T)=-2n_{0}(X)\int^{T} n_{1}(X,T')dT',
\label{r}
\end{equation}
connecting the amplitude $n_{1}$ and the phase $\phi_{1}$.

Proceeding to the next order, namely to order $O(\epsilon^{2})$ and to order 
$O(\epsilon^{5/2})$, Eqs. (\ref{h2}) and (\ref{h1}) respectively read: 
\begin{eqnarray}
&& \phi_{2T}+2n_{0}n_{2}=-n_{1}^{2}-\frac{1}{2}C^{-
2}\phi_{1T}^{2}+\frac{1}{4}C^{-2}n_{0}^{-1}n_{1TT}, \label{nl1} \\
&& n_{2T}+C^{-2}n_{0}\phi_{2TT}=-C^{-2}(n_{1}\phi_{1T})_{T}  \nonumber \\
&&+C^{-1}(n_{0X}\phi_{1T}+2n_{0}\phi_{1XT})+\frac{dC^{-1}}{dX}n_{0}\phi_{1T}. 
\label{nl2}
\end{eqnarray}
The compatibility condition of Eqs. (\ref{nl1}) and (\ref{nl2}) yields the 
equation $1-2n_{0}^{2}C^{-2}=0$ 
for the velocity $C$, 
along with the following nonlinear evolution equation for $n_{1}$,
\begin{equation}
n_{1X}-\frac{C}{n_{0}^{3}}n_{1}n_{1T}+\frac{1}{32n_{0}^{4}C}n_{1TTT}
=-\frac{d}{dX}(\ln n_{0})n_{1},
\label{kdv1}
\end{equation}
which is obtained upon also using Eq. (\ref{r}). Equation (\ref{kdv1}) has 
the form of a KdV equation with variable coefficients, which has been 
used to describe shallow water-waves over variable depth, or ion-acoustic 
solitons in inhomogeneous plasmas \cite{asano}. 
The inhomogeneity-induced dynamics and dissipation of the KdV solitons 
has been studied analytically \cite{ko,karpman}. Below, we will employ 
these results to 
analyze the coherent evolution and dissipation of dark solitons in an atomic gas in the TG limit.

Introducing the transformations $\chi=\int(32n_{0}^{4}C)^{-1}dX$, 
$\tau=T$ and $n_{1}=(3/32n_{0}^{3})\upsilon(\chi,\tau)$, we put Eq. 
(\ref{kdv1}) into the form,
\begin{equation}
\upsilon_{\chi}-6\upsilon\upsilon_{\tau}+\upsilon_{\tau \tau 
\tau}=\lambda(\chi)\upsilon,
\label{kdv2}
\end{equation}
where $\lambda(\chi) \equiv 2(\ln n_{0})_{\chi}$. In the case $\lambda=0$, i.e., 
for a homogeneous gas with $n_{0}(X)=n_{\rm o}={\rm const.}$, Eq. (\ref{kdv2}) is the 
completely integrable KdV equation, which possesses a single-soliton solution of 
the following form 
\cite{abl},
\begin{equation}
\upsilon=-2\kappa^{2}{\rm sech}^{2}Z, \,\,\ Z=\kappa\left[\tau-
\zeta(\chi)\right],
\label{sol}
\end{equation}
where $\zeta(\chi)=4\kappa^{2}\chi+\zeta_{0}$ is the soliton center (with
$d\zeta/d\chi=4\kappa^{2}$ being the soliton velocity in the $\tau$-$\chi$ 
reference frame),
while $\kappa$ and $\zeta_{0}$ are arbitrary constants presenting the soliton's 
amplitude (as well as inverse temporal width) and initial position respectively. 
Apparently, 
Eq. (\ref{sol}) describes a density notch on the backround density $n_{\rm o}$, with 
a phase jump across it [see Eq. (\ref{r}), which implies that $\phi_{1}\sim 
\tanh Z$] and, thus, it represents an approximate dark soliton solution of Eq. 
(\ref{gpe}).

In the general case of the inhomogeneous gas (i.e., in the 
presence of the trapping potential), soliton dynamics can still be studied 
analytically, provided that the
right-hand side of Eq. (\ref{kdv2}) can be treated as a small perturbation. As 
$\lambda(\chi)$ is directly proportional to the density gradient, such a 
perturbative study is relevant in regions of small density gradients, 
which is consistent with the use of the local density approximation.
In this case, employing the perturbation theory for solitons \cite{ps}, we  
express the solution of Eq. (\ref{kdv2}) as, 
\begin{equation}
\upsilon=\upsilon_{S}+\upsilon_{R},
\label{spr}
\end{equation}
where $\upsilon_{S}$ is the soliton part, which has the same functional form as 
in the unperturbed homogeneous case (c.f. Eq. (\ref{sol})), but with the soliton 
parameters $\kappa$ and $\zeta$ being now unknown functions of $\chi$. The 
contribution $\upsilon_{R}$, being of the same order of smallness as $\lambda$, 
denotes the radiation part of the solution (i.e., the sound profile) due to the 
effect of axial inhomogeneity. Following \cite{karpman}, we first derive the 
following evolution equations for the soliton's amplitude $\kappa(\chi)$ and 
center $\zeta(\chi)$,
\begin{eqnarray}
\frac{d\kappa}{d\chi}=\frac{2}{3}\kappa\lambda, \,\,\,\,\,\
\frac{d\zeta}{d\chi}=4\kappa^{2}+\frac{\lambda}{3\kappa}. \label{ev}
\end{eqnarray}
These equations can be solved analytically and the results, expressed in terms 
of the slow variable $X$, read:
\begin{eqnarray}
&&\kappa(X)=\kappa(0)\left(\frac{n_{0}(X)}{n_{0}(0)}\right)^{4/3}, \label{kx1} 
\\ 
&&\zeta(X)=\frac{\kappa^{2}(0)}{8\sqrt{2}s n_{0}^{8/3}(0)}\int_{0}^{X} n_{0}^{-
7/3}(X')dX' \nonumber \\
&&+\frac{1}{3\kappa(0)}\left[1-3\left(\frac{n_{0}(X)}{n_{0}(0)}\right)^{-
1/3}\right],
\label{kx2}
\end{eqnarray}
where $\kappa(0)$ and $n_{0}(0)$ are respectively the soliton amplitude and 
density at $X=0$, while $s=\pm 1$ represents the two possible directions of the 
soliton propagation. Additionally, we find the following approximate (for 
$|Z|\gg 1$) expression for the
radiation part of the solution (i.e., sound wave emitted by the soliton),
\begin{equation}
\upsilon_{R}\approx -\frac{32\sqrt{2}s 
n_{0}^{4/3}(0)}{3\kappa(0)}n_{0}^{8/3}(X)\frac{dn_{0}(X)}{dX}(1-\tanh Z),
\label{radg}
\end{equation}
where $Z=\kappa(X)\left[\tau-\zeta(X)\right]$ is the soliton phase,
with $\kappa(X)$ and $\zeta(X)$ given by Eqs. (\ref{kx1})-(\ref{kx2}).

Based on the above results for the evolution of the soliton parameters, we will 
now show that the dark soliton will display an oscillatory motion in the 
harmonic trap 
$V(X)=(1/2)\Omega^{2}X^{2}$. This can be done upon deriving  
the phase of the soliton, which, to order $O(\epsilon^{3/2})$, reads
\begin{eqnarray}
&&Z=\epsilon^{1/2}\mu_{0}^{2/3}\kappa (0) \left[1-\frac{4}{3} \left( \frac{X}{L} 
\right)^{2} \right] \nonumber \\
&&\times \left[t-\int \frac{dX}{C}-\epsilon\frac{\kappa^{2}(0) \mu_{0}^{-
2/3}}{8s\Omega}\left( \frac{X}{L} \right)\right]
\label{solphase}
\end{eqnarray}
(recall that $L=\sqrt{2\mu_{0}}/\Omega$ defines the axial size of the gas). 
Then, looking along the characteristic lines of soliton motion, it is easy to 
show that the position of the soliton satisfies the following equation of 
motion,
\begin{equation}
\frac{dX}{dt}=\frac{8\sqrt{2}s\Omega L n_{0}(X)}{8\Omega L+\epsilon \sqrt{2} 
\kappa^{2}(0)\mu_{0}^{-2/3}n_{0}(X)}.
\label{solpos}
\end{equation}
For sufficiently small $\epsilon$ the second term in the denominator can be neglected and the separable resulting equation can readily be integrated. In particular, taking into regard Eq. (\ref{TF}) for a parabolic trap, we find that 
\begin{equation}
X=L\sin(\Omega t),
\label{m}
\end{equation}
which demonstrates that a shallow dark soliton in the TG limit oscillates at the trap frequency.

In conclusion, we have developed a systematic analytical approach, based on a 
reductive 
perturbation method, to study the linear and nonlinear excitations of a 
Bose gas of impenetrable bosons (i.e., in the Tonks-Girardeau limit). We have recovered the 
Bogoliubov spectrum of linear excitations, with excitation frequencies (and 
speed of sound) varying slowly along the axial direction. Additionally, we have 
shown that shallow dark solitons can be 
described by an effective Korteweg-de Vries equation with variable coefficients. 
We have found analytical expressions for the 
inhomogeneity-induced evolution of the soliton parameters (amplitude, width, 
position, velocity) and the profile of the sound emitted by the soliton. 

Our results are based on the quintic nonlinear Schr{\"o}dinger equation 
\cite{kolom1,bhad,kolom2,tanatar,kim,lee0,lieb,anna}, 
which is expected to be valid for weak density modulations. This approach 
enables {\em analytical} results to be obtained, and the oscillation frequency of shallow solitons is thus found to be identical to 
the one obtained via numerical simulations based on the Bose-Fermi mapping 
\cite{busch}. 
This result seems to additionally justify {\it a posteriori} the use of 
Eq. (\ref{gpetg}) for shallow solitons. In particular, our work demonstrates the validity of the reductive perturbation method for dynamics of shallow solitons in the Tonks-Girardeau (TG) regime, while an earlier study \cite{huang} confirmed its usefulness in the opposite regime of  Bose-Einstein condensation (BEC). The crossover between these two regimes \cite{dunjko} is important, and diagnostics (e.g. \cite{chiara}) for the degree of ``impenetrability'' of a 
trapped one-dimensional Bose gas are required for interpreting current and  
future experiments \cite{Tonks_Exp}. 
The presented technique paves the way 
for investigating the crossover in the 
oscillation frequency of shallow dark solitons [ranging from $\Omega/\sqrt{2}$ 
in the 
BEC regime, to $\Omega$ in the 
TG regime ($\Omega$ is the trap frequency)], upon suitably generalizing the nonlinearity  to a smooth function interpolating between 
these two regimes \cite{dunjko,brand}.



We acknowledge fruitful discussions with J. Brand and L. Santos.
This work was supported by 
the Eppley Foundation for Research, NSF-DMS-0204585 and NSF-CAREER (PGK).


\end{multicols}


\begin{references}

\vspace{-15mm}


\bibitem{dark} S. Burger {\it et al.}, Phys.\ Rev.\ Lett.\ \textbf{83}, 5198 (1999); 
J. Denschlag {\it et al.} Science \textbf{287}, 97 (2000);
B.P.\ Anderson {\it et al.}, Phys.\ Rev.\ Lett.\ \textbf{86}, 2926 (2001);
Z. Dutton {\it et al.}, Science \textbf{293}, 663 (2001).

\bibitem{motion} 
Th.\ Busch and J.R.\ Anglin, Phys.\ Rev.\ Lett.\ \textbf{84}, 2298 (2000).

\bibitem{huang} G. Huang {\it et al.}, 
\newblock Phys.\ Rev.\ A 
\textbf{65}, 053605 (2002).

\bibitem{frantz} D.J.\ Frantzeskakis {\it et al.} \newblock Phys.\ Rev.\ A 
\textbf{66}, 053608 (2002).

\bibitem{dis} P.O.\ Fedichev {\it et al.},  
\newblock Phys.\ Rev.\ A \textbf{60}, 3220 (1999); A.\ Muryshev {\it et al.},  
Phys.\ Rev.\ Lett.\ \textbf{89}, 110401 (2002).

\bibitem{thsnbec} D.L.\ Feder {\it et al.}, Phys.\ Rev.\ A \textbf{62}, 053606  
(2000); J.\ Brand and W.P.\ Reinhardt, Phys.\ Rev.\ A \textbf{65}, 043612 (2002).

\bibitem{nick} N.G.\ Parker {\it et al.}, \newblock Phys.\ Rev.\ Lett.\ 
\textbf{90}, 220401 (2003).

\bibitem{nick2} N.G.\ Parker {\it et al.}, \newblock J. Phys. B {\bf 36}, 2891 
(2003).

\bibitem{george} G.\ Theocharis {\it et al.}, Phys.\ Rev.\ Lett.\ \textbf{90}, 
120403 (2003).

%

\bibitem{ols1} M. Olshanii, Phys.\ Rev.\ Lett.\ \textbf{81}, 938 (1998).

\bibitem{tg} L. Tonks, Phys. Rev. {\bf 50}, 955 (1936);
M. Girardeau, J. Math. Phys. (N.Y.) {\bf 1}, 516 (1960).

\bibitem{Tonks_Exp} H. Moritz {\it et al.},  Phys.\ Rev.\ Lett.\ \textbf{91}, 250402 (2003); 
J. Reichel and J.H. Thywissen, cond-mat/0310330 (2003); 
B. Laburthe {\it et al.}, cond-mat/0312003 (2003).

\bibitem{LL} E.H. Lieb and W. Liniger, Phys.\ Rev.\ \textbf{130}, 1605 (1963).

\bibitem{gw} M.D. Girardeau and E.M. Wright, Phys.\ Rev.\ Lett.\ \textbf{84}, 
5691 (2000).

\bibitem{gw3} M.D. Girardeau, Phys. Rev. Lett. {\bf 91}, 040401 (2003).

\bibitem{petrov} D.S. Petrov {\it et al.},  
Phys.\ Rev.\ Lett.\ \textbf{85}, 3745 (2000).

\bibitem{dunjko} V. Dunjko {\it et al.},
Phys.\ Rev.\ Lett.\ \textbf{86}, 5413 (2001).


\bibitem{santos} P. \"{O}hberg and L. Santos, Phys.\ Rev.\ Lett.\ \textbf{89}, 
240402 (2002).



\bibitem{busch} Th. Busch and G. Huyet, J. Phys. B {\bf 36}, 2553 (2003).

\bibitem{kolom1} E.B. Kolomeisky {\it et al.}, Phys. Rev. Lett. {\bf 85}, 1146 
(2000).

\bibitem{bhad} R.K. Badhuri {\it et al} J. Phys. A {\bf 34}, 6553 (2001).

\bibitem{fermi} B. Damski {\it et al}, J. Phys. B {\bf 35}, L153 (2002);
T. Karpiuk {\it et al}, J. Phys. B {\bf 35}, L315 (2002).

\bibitem{review} F. Dalfovo {\it et al.} Rev. Mod. Phys. {\bf 71}, 463 (1999).

\bibitem{kolom2} E.B. Kolomeisky and J.P. Straley, Phys. Rev. B {\bf 46}, 11749 
(1992);

\bibitem{tanatar} B. Tanatar, Europhys. Lett. {\bf 51}, 261 (2000).

\bibitem{kim} Y.E. Kim and A.L. Zubarev, Phys. Rev. A {\bf 67}, 015602 (2003).

\bibitem{lee0} M.D. Lee {\it et al.}, 
Phys. Rev. A {\bf 65}, 043617 (2002); 
M.D. Lee et al., 
cond-mat/0305416 (2003).

\bibitem{lieb} E.H. Lieb {\it et al.}, 
Phys. Rev. Lett. {\bf 91}, 150401 (2003).

\bibitem{anna} A. Minguzzi {\it et al.}, Phys. Rev. A {\bf 64}, 033605 (2001).

\bibitem{brand} J. Brand, cond-mat/0311206 (2003).

\bibitem{gw2} M.D. Girardeau and E.M. Wright, Phys.\ Rev.\ Lett.\ \textbf{84}, 
5239 (2000).


\bibitem{chiara} C. Menotti and S. Stringari, Phys. Rev. A {\bf 66}, 043610 
(2002).



\bibitem{landau} L.D. Landau and E.M. Lifshitz, {\it Fluid Mechanics} (Pergamon, New York, 1959).

\bibitem{rpm} T. Taniuti, Prog. Theor. Phys. Suppl. {\bf 55}, 1 (1974).

\bibitem{asano} N. Asano, Prog. Theor. Phys. Suppl. {\bf 55}, 52 (1974).

\bibitem{ko} K. Ko and H.H. Kuehl, Phys. Rev. Lett. {\bf 40}, 233 (1978);
K. Ko and H.H. Kuehl, Phys. Fluids {\bf 23}, 834 (1980).

\bibitem{karpman} V.I. Karpman and E.M. Maslov, Phys. Fluids {\bf 25}, 1686 (1982).

\bibitem{abl} M.J. Ablowitz and P.A. Clarkson, {\it Solitons, Nonlinear Evolution Equations and Inverse Scattering} (Cambridge University Press, Cambridge, England, 1991).

\bibitem{ps} V.I. Karpman and E.M. Maslov, Zh. Eksp. Teor. Fiz. {\bf 75}, 504 (1978) [Sov. Phys.-JETP {\bf 48}, 252 (1978)]; V.I. Karpman, Phys. Scripta {\bf 20}, 462 (1979).





\end{references}
\end{document}